# A VISUAL STUDY ON THE SPRAY OF GAS-LIQUID ATOMIZER

Maziar Shafaee[1*], Mohammad Hosein Sabour[2], Armin Abdehkakha[3], Abbas Elkaie[4]

[1]Proffessor Assistant, Faculty of New Sciences and Technologies, Tehran University, mshafaee@ut.ac.ir

[2]Proffessor Assistant, Faculty of New Sciences and Technologies, Tehran University, saburmh@ut.ac.ir

[3]Master student, Faculty of New Sciences and Technologies, Tehran University, arminabdeh@ut.ac.ir

[4]Master student, Faculty of New Sciences and Technologies, Tehran University, a.elkaie@ut.ac.ir

*(Phone: +989190110200)*

**Abstract.** A Visual investigation of spray cone angle for different air-blast atomizers and flow conditions is described. Liquid jets are exposed to high gas stream with specific relative angle. Using high speed camera, spray cone angle over a range of Reynolds number $4 \times 10^4$ to $9 \times 10^4$ and Weber number 1 to 140 is studied, followed by laser-based diagnosis of particle distribution and Sauter mean diameter. The results show that for high Reynolds and Weber number, the cone angle is independent of flow condition, and is only dependent on geometry of atomizer, including orifice diameter with significant effect on cone angle, Sauter mean diameter and particle distribution.

## 1. Introduction

Two-fluid atomization (also termed as twin-fluid, two-phase, pneumatic and aerodynamic atomization) is one of the liquid disintegration techniques applied to various spraying systems (Shafaee et al., 2014). This type of atomization may be divided into two categories including air-assist and air-blast atomization. In both processes, the bulk liquid to be atomized is first transformed into a jet or sheet at a relatively low velocity and then exposed to a high velocity gas stream (Lefebvre, 1992). The kinetic energy of the gas flow is used as a source of atomization to shatter the bulk liquid into ligaments that subsequently breakup into droplets (Marmottant et al., 2004 & Sankarakrishnan et al., 2008).

The penetration, spray dispersion angle and droplet sizes related to the breakup process for liquid jets and air/fuel distributions are very important parameters in propulsion systems requiring combustion efficiency and regulation of pollutant emissions (Lee et al., 2010). Spray angle is also one of the important external spray characteristics for evaluating the atomizers performance. Most of the sprays have a conical shape wherein the cone angle is usually defined as the angle between the tangents to the spray envelope at the atomizer exit. The spray angle of a two-fluid atomizer should be such that it could provide a good mixing between the two fluids which cause the liquid jets to be disintegrated perfectly through the gas stream. In combustion systems, the value to be selected for the cone angle will depend on the shape of the combustion chamber prior to the air and fuel mixing conditions.

Chatterjee et al. (Chatterjee et al., 2004) investigated the effect of spray cone angle on combustion performance of a liquid fuel spray in a gas turbine combustor. They observed that an increase in spray cone angle increases the wall temperature. Therefore, it is very important to develop an accurate method for predicting the spray cone angle in such atomizers.

Guo et al. (Guo et al., 2002) investigated two-phase spraying characteristics of a gas-liquid nozzle used for the humidification of smoke. They found that at the given gas pressure the spray angle will increase gradually with the increasing of liquid phase velocity, while at the given liquid pressure the spray angle decreases with increasing the gas pressure.

Varde (Varde, 1985) made a liquid fuel spray injected into a gaseous environment in order to investigate the effects of nozzle orifice size and operating conditions on the spray cone angle. The results showed that the spray cone angle is dependent on the orifice dimensions as well as on the nozzle operating conditions. In addition, he derived a correlation to predict spray cone angle in terms of Reynolds and Weber numbers.

Chen et al. (Chen et al,. 1992) measured the spray cone angle of a pressure swirl atomizer for five different values of discharge orifice length/diameter ratio ($l_0/d_0$). They found that an increase in $l_0/d_0$ always leads to a reduction in spray cone angle.

Reitz and Bracco (Bracco, 1979) proposed a correlation for estimating the spray cone angle versus nozzle length and nozzle hole diameter. Also, Hiroyasu and Arai (Hiroyasu et al., 1980) established a correlation between the spray cone angle and the nozzle operational conditions and its geometrical parameters.

1. **Experimental setup**

*Figure 1* shows a schematic of the experimental setup used in this study comprising three main parts, i.e. the liquid (water) feed line, the compressed gas (air) line and the injector, in addition to the configuration of a high speed video camera and Malvern Master Sizer X laser systems.

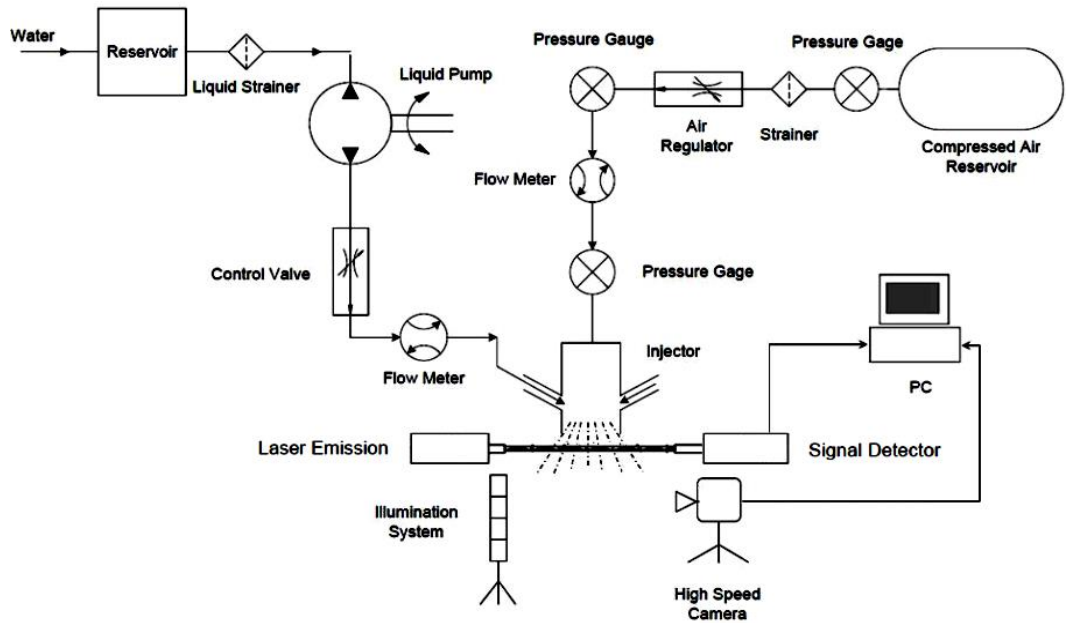

*Figure 1. Schematic view of the experimental setup.*

The first part, the liquid feed line, consists of five elements including a liquid reservoir with a capacity of 1.1 m³ connected to the main tap water, a stainless steel mesh strainer hampering any possible tiny debris from the liquid flow, a liquid piston pump with a regulating pressure in the range of 0 to 50 bar capable of providing liquid flowrates up to 50 L/min, a needle valve for flowrate adjustment and finally a two-fluid atomizer and related fixture.

The second part, the compressed air line, is composed of two elements including a pre-charged compressed air reservoir having a capacity of 50 L with a maximum allowable pressure of 140 bar and an air pressure regulator. The third part includes an injector, i.e. a two-fluid atomizer, which is connected to the water and compressed air lines using an interface fixture.

A schematic sketch of the atomizer and its picture are shown in *Figure 2*. The compressed air flows through the central part of the atomizer while the liquid feeds through an annular passage which is placed with an angle of 55° relative to the atomizer central axis and mounted on the atomizer periphery, to intercept with the compressed air flow. The geometrical parameters of the atomizer, with reference to *Figure 2a*, include the liquid inlet diameter $d_l$=1.6 mm, the atomizer exit diameter $d_g$=21.5 mm and the air and liquid mixing length $l$ =8mm. But the geometrical parameters of the investigated atomizers are shown in *Table 1*.

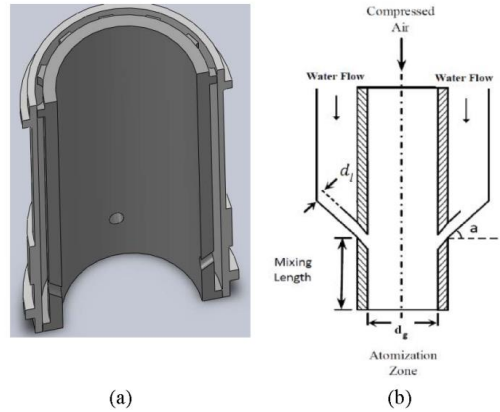

*Figure 2. The two-fluid atomizer used in this study, (a) 3-D view, (b) a schematic.*

*Table 1. Geometrical specifications of the investigated two-fluid atomizers.*

| No | 1 | 2 | 3 | 4 | 5 | 6 | 7 | 8 | 9 | 10 | 11 | 12 | 13 | 14 | 15 | 16 | 17 | 18 | 19 | 20 | 21 | 22 | 23 | 24 | 25 | 26 | 27 | 28 |
|---|---|---|---|---|---|---|---|---|---|---|---|---|---|---|---|---|---|---|---|---|---|---|---|---|---|---|---|---|
| $d_l$ (mm) | 1.0 | | | | | | | 1.2 | | | | | | | 1.6 | | | | | | | 1.8 | | | | | | |
| $a$ (°) | 45 | | | | 40 | 35 | 30 | 45 | | | | 40 | 35 | 30 | 45 | | | 40 | 35 | 30 | | 45 | | | | 40 | 35 | 30 |
| $l$ (mm) | 6 | 8 | 10 | 12 | 8 | 8 | 8 | 6 | 8 | 10 | 12 | 8 | 8 | 8 | 6 | 8 | 10 | 12 | 8 | 8 | 8 | 6 | 8 | 10 | 12 | 8 | 8 | 8 |

The visualization system employed in this study consists of a high speed digital camera (Mega Speed MS50K) set at a recording rate of 2500 fps which is capable of recording image files with a resolution of 640×480 pixels and simultaneously transmitted the captured files to a dedicated PC. The spray angle of each atomizer has been measured by applying an image processing method through a frame by frame analysis *(Fig. 3)*. Also, The droplet sizes and their distribution have been measured using Malvern Master Sizer X.

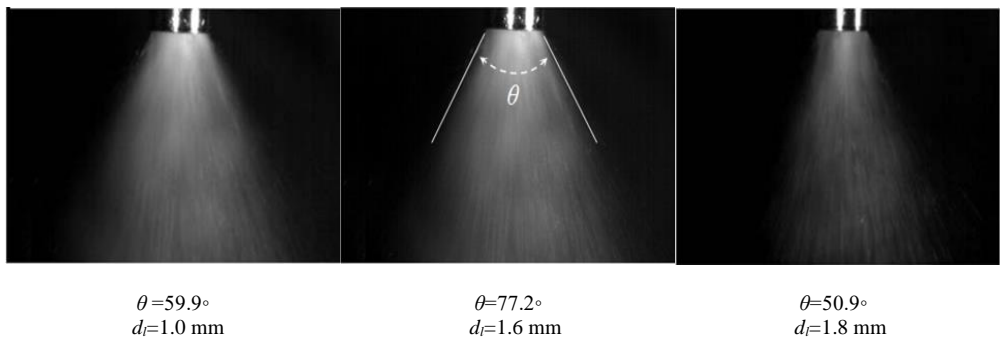

$\theta$ =59.9°  
$d_l$=1.0 mm

$\theta$=77.2°  
$d_l$=1.6 mm

$\theta$=50.9°  
$d_l$=1.8 mm

*Figure 3. Spray cone angle in different liquid port diameters (a=45°, l=6 mm, $Re_l$=90000, $We_g$=140).*

## 2. Results and Discussion

In the present atomizer, the transverse interaction between the gas stream (air) and each of the liquid jets (six liquid jets), causes a complicated atomization phenomenon (Shafaee et al., 2014). A standard dimensional analysis (Buckingham's p-theorem) shows that all the parameters affect the spray cone angle can be classified into two main groups including geometrical parameters and operating conditions (*Eq. 1*).

$$\theta = f(d_g,\ d_l,\ l,\ a,\ u_g,\ u_l,\ \rho_g,\ \rho_l,\ \mu_g,\ \mu_l,\ \sigma) \qquad (1)$$

Applying dimensional analysis leads to calculate the angle in terms of dimensionless groups (*Eq. 2*).

$$\theta = f(\frac{d_g}{d_l},\frac{l}{d_l},a,\frac{\rho_g u_g}{\rho_l u_l},\text{Re}_l,We_g) \qquad (2)$$

In which $\text{Re}_l = \frac{\rho_l u_l d_l}{\mu_l}$ and $We_g = \frac{\rho_g (u_g - u_l)^2 d_l}{\sigma}$.

First, the effect of flow conditions on spray angle at a fixed geometry of atomizer is investigated. Then, the effect of geometric parameters on the angle is studied. Visual investigation shows that increasing the operational parameters i.e. Reynolds and Weber numbers in a constant geometry causes the atomizer to pass four different breakup regimes including Rayleigh, first wind induced, second wind induced and atomization mode. These regimes are introduced in (Lin et al., 1998).

In Rayleigh mode *(Fig. 4a)*, the size of droplets are greater than the jet diameter (liquid port diameter), while in first wind induced regime, the droplet sizes are in order of the jet diameter *(Fig. 4b)*. *Figure 4c* shows the second wind induced regime with droplets approximately ranging from 100μ to 1450μ. The final mode, atomization, has fine droplets with *SMD*=55μ (*We_g*=49.8) and approaches to *SMD*=34μ as Weber reaches to 150 and varies little with Weber number in higher values of *We_g*.

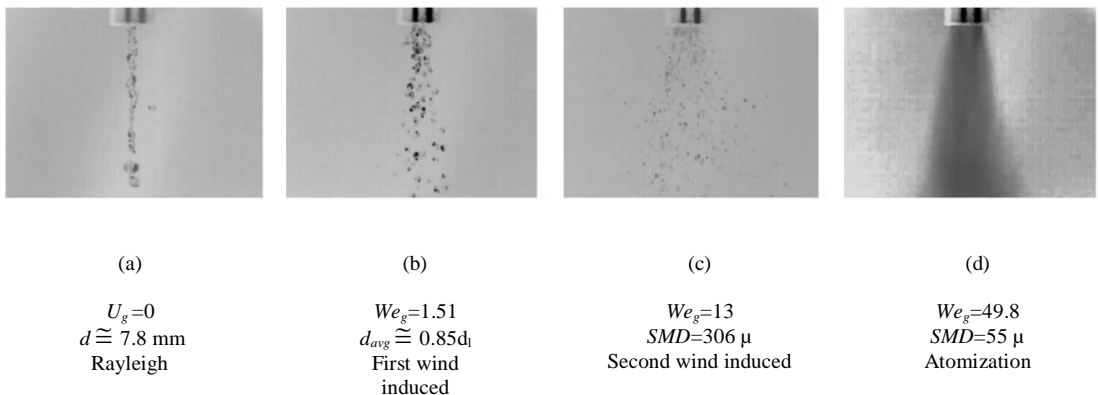

(a)      (b)      (c)      (d)

$U_g$=0      $We_g$=1.51      $We_g$=13      $We_g$=49.8
$d \cong$ 7.8 mm      $d_{avg} \cong 0.85 d_l$      SMD=306 μ      SMD=55 μ
Rayleigh      First wind induced      Second wind induced      Atomization

**Figure 4.** *Different breakup modes based on Weber increment ($Re_l$ =1650).*

Spray size distribution obtained by Malvern Master Sizer X is shown in *Figure 5*. The results show that the spray size distribution is affected by atomizing air Weber number. The curves show that as Weber number increases, the size distribution curve moves to the small sizes.

*Figure 6* shows the spray cone angle produced by the two-fluid atomizer as a function of its operating conditions. The curves show that in a constant geometric of atomizer, a similar trend is seen for spray angle variations. In each Weber number, an increase in Reynolds number causes a decrease in spray angle followed by approaching to an asymptotic value for higher Reynolds numbers. Also, increase in the Weber number for a constant value of Reynolds causes the spray cone angle to decrease. The rate of this decrease is considerable at low Weber numbers while for higher values of Weber, the curves will coincide with each other. It means that in constant Reynolds and high values of Weber, a further increase in Weber number does not decrease the spray angle. So, it can be concluded that for $We_g$>140, the spray angle becomes less dependent to Weber number. Consequently, in the present two-fluid atomizer, there are ranges of operating conditions in which the spray angle varies little with Reynolds and Weber numbers. In these ranges, the spray cone angle may only depend to the geometrical parameters of the atomizer.

The following empirical correlation has been developed applying a multiple variable least square regression technique on a set of 36 experimental data to predict the spray cone angle as a function of its operating conditions, i.e. Reynolds and Weber numbers:

$$\theta = -12.56\ln\left(\frac{Re_l}{10^4}\right) - 1.55\ln\left(\frac{We_g}{10}\right) + 97.72 \qquad (3)$$

$$4\times 10^4 \leq Re_l \leq 9\times 10^4, We_g \leq 140$$

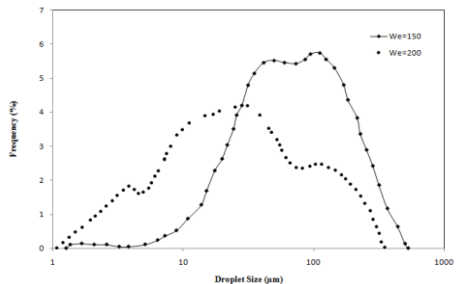

**Figure 5.** *Droplet size distribution based on Weber increment ($Re_l$ =1650).*

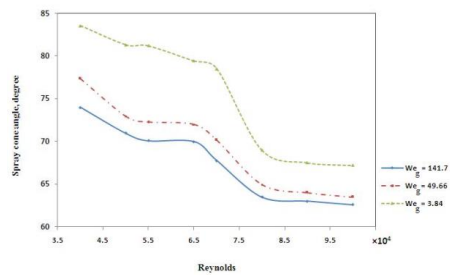

**Figure 6.** *Spray angle variations against Reynolds in different Weber numbers.*

As mentioned above, there are ranges of operating conditions in which the spray angle varies little with Reynolds and Weber numbers ($We_g$>140, $Re_l$>85000). Therefore, in these ranges of operational conditions, the spray cone angle may only depend to the geometrical parameters of the atomizer (Eq. (4)).

$$\theta = f(\frac{d_g}{d_l}, \frac{l}{d_l}, a) \ , \ Re_l > 85000, We_g > 140 \tag{3}$$

It should be noted that since $d_g$ is constant, thus $\theta$ is a function of $d_l$, $l$ and $a$. In the following, the effect of atomizer geometry on spray angle in the range where flow condition has no significant effect on spray angle has been studied.

*Figure 7* shows the effect of geometrical parameters on the droplet sizes in atomization mode. Decreasing the liquid jet diameter cause the atomizer to generate more fine droplets with smaller SMD. The same trend with a less intensity is also observed for mixing length of the atomizer. However, any change in the injection angle of the liquid jets, cause a slight effect on the droplets sizes. So, it is clear that the liquid jet diameter has the greatest effect on the droplets sizes.

As *Figure 7* shows, in all geometrical parameters, increasing the Weber number causes the *SMD* to decrease. The curves show that at higher Weber numbers, the effect of $We_g$ on droplet size decreasing become negligible.

In order to reduce the errors encountered in measurement procedure, the image processing method which is used for spray angle measurement has been repeated 5 times to assure the repeatability of the results obtained for each geometry. *Figure 8* shows the effect of liquid jet diameter on spray cone angle at various mixing lengths. As the liquid jet diameter increases, the spray cone angle increases with descending slope until reaches a maximum value. Then with increase of liquid port diameter, the spray angle begins to decrease. In this type of atomizer, the spray cone angle may depend on two main parameters which are the liquid flow rate with an increasing effect on spray cone angle and the gas jet velocity with an opposite effect. As the liquid port diameter is increased for a constant Reynolds and Weber number, the liquid flow rate should be increased and the gas jet velocity should be decreased. As the liquid port diameter is increased, first, the increasing effect of liquid flow rate on spray angle is dominant and then the decreasing effect of gas jet velocity becomes conqueror.

From *Figure 8*, it is also found that increasing the length of mixing chamber ($l$) causes a little decrease in spray cone angle. Therefore, it can be concluded from *Figure 7* and *Figure 8*, that the liquid jet diameter has a greater effect on both droplet sizes and spray cone angle respect to the other geometrical parameters.

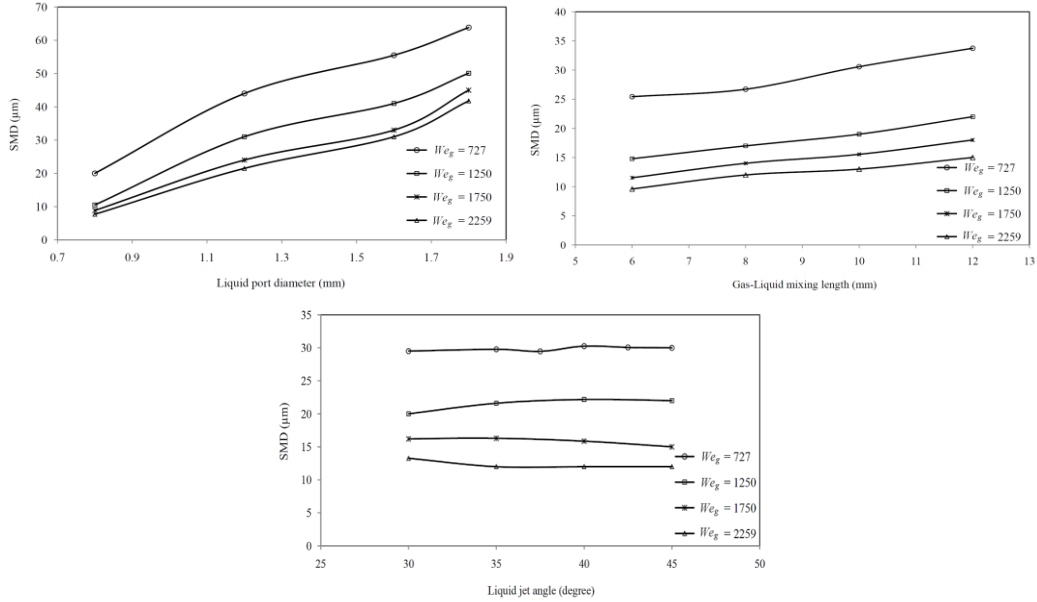

*Figure 7.* Effect of geometrical parameters on the droplets sizes ($Re_l$=1650).

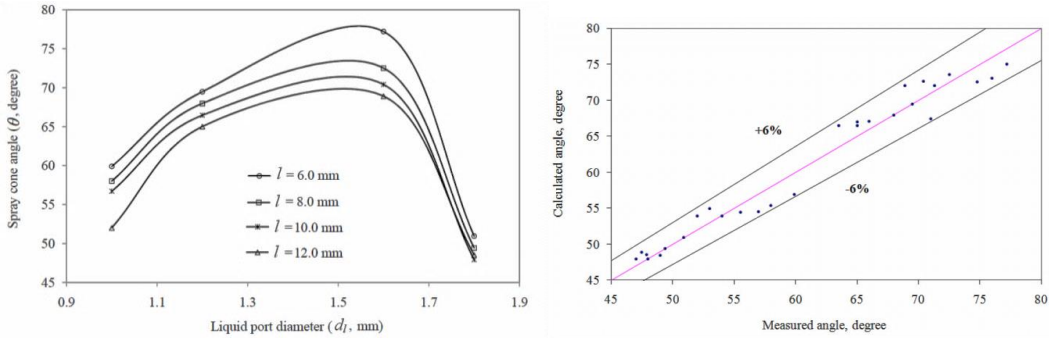

*Figure 8.* Variations of spray cone angle versus the liquid port diameter for different values of mixing lengths ($Re_l$=90000, $We_g$=140).

*Figure 9.* Correlation accuracy plot.

The following empirical correlation has been developed to predict the spray cone angle in terms of geometrical parameters in the ranges that this angle varies little with operational conditions. Applying a multiple variable least square regression technique on a set of 28 experimental data (atomizers listed in *Table 1*), the *Eq. 5* is developed.

$$\theta = 375 + 386 d_l^2 - 261\exp(d_l) + 0.8\frac{a}{l} \tag{5}$$

where $d_l$ and $l$ are in millimeter and $a$ and $\theta$ are in degree. An error analysis indicates that Eq. (5) is applicable over the entire range which is defined for geometrical parameters with a maximum error of 6%. Spray cone angles which are calculated using the above correlation are plotted against the measured values in *Figure 9*. In most cases the correlation is seen to predict the spray cone angle fairly accurately.

## 3. Conclusions

An experimental investigation is conducted to determine the effect of operating condition on the spray cone angle of a two-fluid atomizer. Using a high speed imaging system, the spray cone angle has been determined in different flow and geometrical conditions. Also, the Malvern Master Sizer X laser system was used to measure Sauter mean diameter of droplets and droplet size distribution. The following conclusion is obtained from this study:

- First, the effects of Reynolds and Weber numbers on the spray cone angle were studied. In each Weber number, Reynolds increasing causes a decrease in spray cone angle followed by approaching to an asymptotic value for higher Reynolds numbers. Similarly, an increase in Weber number in a constant value of Reynolds causes the spray cone angle to decrease with a descending slope.
- In the present atomizer, there are ranges of operating conditions in which the Reynolds and Weber numbers have a little effect on spray cone angle and it may only depends to the geometrical parameters of the atomizer.

  The following empirical correlation has also been presented to predict the spray cone angle as a function of its operating conditions. This equation is applicable over the entire range defined for Reynolds and Weber numbers with a maximum error of 6.4%.

$$\theta = -12.56\ln\left(\frac{\text{Re}_l}{10^4}\right) - 1.55\ln\left(\frac{We_g}{10}\right) + 97.72$$
$$4 \times 10^4 \leq \text{Re}_l \leq 9 \times 10^4, We_g \leq 140$$

- In the present atomizer, the mixing length and injection angle have no significant effect on spray cone angle.

- At constant Reynolds and Weber numbers, from ranges in which their effect on spray cone angle is not considerable, liquid jet diameter is dominant among all geometrical parameters.
- The following correlation has been peresented to predict the spray cone angle of two-fluid atomizer in terms of geometrical parameters.

$$\theta = 375 + 386 d_l^2 - 261 \exp(d_l) + 0.8 \frac{a}{l}$$